\documentclass[12pt,a4paper]{article}

\usepackage[margin=1in]{geometry}
\usepackage{amsmath,amssymb}
\usepackage{graphicx}
\usepackage{booktabs}
\usepackage{array}
\usepackage{multirow}
\usepackage{longtable}
\usepackage{tabularx}
\usepackage{caption}
\usepackage{float}
\usepackage{natbib}
\usepackage{hyperref}
\usepackage{url}
\usepackage{xcolor}
\usepackage{setspace}
\usepackage{titlesec}
\usepackage{enumitem}
\usepackage{parskip}
\usepackage{rotating}
\usepackage{adjustbox}
\usepackage{mdframed}

\graphicspath{{./}}

\hypersetup{
  colorlinks = true,
  linkcolor  = blue!60!black,
  citecolor  = blue!60!black,
  urlcolor   = blue!60!black
}

\titleformat{\section}{\large\bfseries}{\thesection}{1em}{}
\titleformat{\subsection}{\normalsize\bfseries}{\thesubsection}{1em}{}
\titleformat{\subsubsection}{\normalsize\itshape}{\thesubsubsection}{1em}{}

\newcommand{\tablenote}[1]{\smallskip\noindent\footnotesize\textit{Note.} #1}

\begin{document}

% -------------------------------------------------------
% TITLE BLOCK
% -------------------------------------------------------
\begin{center}
  {\LARGE\bfseries Construct Validity of AI-Administered Cognitive Assessment:\\[4pt]
  Psychometric Validation of a Nine-Dimension Process-Oriented\\[4pt]
  Evaluation Instrument}\\[18pt]
  {\large Abdalla Doleh$^{1,*}$ \quad Toni Somers$^{2}$ \quad Ratna Babu Chinnam$^{1}$}\\[8pt]
  {\normalsize
  $^{1}$Department of Industrial \& Systems Engineering,
  Wayne State University, Detroit, MI, USA\\[3pt]
  $^{2}$Dean's Office, Mike Ilitch School of Business,
  Wayne State University, Detroit, MI, USA\\[6pt]
  $^{*}$Corresponding author: \href{mailto:ai5145@wayne.edu}{ai5145@wayne.edu}\quad
  ORCID: \href{https://orcid.org/0009-0008-5192-2167}{0009-0008-5192-2167}\\[2pt]
  Ratna Babu Chinnam: \href{mailto:ai2396@wayne.edu}{ai2396@wayne.edu}}\\[10pt]
  {\normalsize\textbf{Preprint DOI:}
  \href{https://doi.org/10.5281/zenodo.20114584}{https://doi.org/10.5281/zenodo.20114584}}\\[4pt]
  {\normalsize\textit{This work is currently in the process of being submitted for peer review.}}
\end{center}

\vspace{10pt}\hrule\vspace{12pt}

% -------------------------------------------------------
% ABSTRACT
% -------------------------------------------------------
\begin{abstract}
Process-oriented evaluation of large language models requires psychometrically defensible
instruments, yet the measurement tools used in AI cognitive assessment have received little
formal validation. This paper reports the construct validation of a 54-item, nine-dimension
instrument that operationalizes the Multi-Dimensional Assessment for AI Cognition (MAAC)
framework through a repeated six-component AI-administered adjudication architecture. The
validation program combined conceptual construct audit, schema-integrity testing, confirmatory
factor analysis, and targeted item revision across multiple instrument versions, with each
version validated on a fresh cohort of approximately 1,800 to 2,200 scenario-response pairs.

The current best instrument state (v4.7, $n = 1{,}774$) achieves acceptable to good
psychometric properties across all nine cognitive dimensions. All nine dimensions meet the
internal consistency threshold ($\alpha \geq 0.70$; range: 0.770--0.922). Confirmatory factor
analysis shows four dimensions at Good fit (CFI $\geq 0.95$) and four at Acceptable fit
(CFI $\geq 0.90$), with 53 of 54 items (98.1\%) loading $\geq 0.40$ on their parent factor.
Schema enforcement produces perfect structural integrity in the finalized validation pipeline.
One item, Hallucination Control~Q6, remains below the loading threshold but improved materially
under the v4.7 \texttt{evidential\_calibration} redesign ($\lambda = 0.315$), surpassing prior
versions without reaching the target threshold. The instrument is therefore publishable as a
largely validated assessment architecture with one documented sub-threshold component. The
validation program also yields a five-category taxonomy of item-level failure modes in
AI-administered instruments---ceiling/floor compression, construct drift, proxy-formula failure,
condition confounding, and fundamental construct mismatch---that may generalize to other AI
evaluation contexts.
\end{abstract}

\noindent\textbf{Keywords:} AI evaluation, construct validity, psychometrics, instrument
validation, large language models, cognitive assessment, measurement science, confirmatory
factor analysis, AI adjudication

\vspace{8pt}\hrule\vspace{12pt}

% -------------------------------------------------------
% 1. INTRODUCTION
% -------------------------------------------------------
\section{Introduction}
\label{sec:intro}

\subsection{The Problem of Measurement in AI Cognitive Assessment}
\label{sec:problem}

Evaluating what an artificial intelligence system can do is one challenge. Evaluating how it
does it is another. Contemporary AI benchmarking has largely addressed the first challenge
through large-scale task suites such as MMLU \citep{Hendrycks2021}, BIG-bench
\citep{Srivastava2022}, and HELM \citep{Liang2022}, producing reliable aggregate accuracy
estimates but offering limited insight into the cognitive processes that generate those outcomes.
Process-oriented evaluation frameworks have emerged as a complement to outcome-based benchmarks,
but they face a deeper challenge: the measurement instruments through which they are administered
must themselves be valid.

This is not a novel problem for measurement science. In psychometrics, the lesson has been
established for decades that a well-motivated theoretical construct does not automatically produce
a working measurement instrument \citep{Messick1995, DeVellis2017, Nunnally1994}. Items must
cohere. Instruments must behave predictably. Constructs must be defined clearly enough that
independent operationalizations converge. When the instrument fails any of these requirements---
even when the underlying construct is theoretically sound---measurement is compromised and
downstream inferences are uninterpretable \citep{Campbell1959, Mokkink2010}.

This paper applies that lesson to a novel setting: AI-based adjudication instruments for
cognitive assessment of large language models. Specifically, it validates the measurement
instrument used to score each dimension of the Multi-Dimensional Assessment for AI Cognition
(MAAC) framework \citep{DolehP2}. The MAAC adjudication instrument is not a simple scoring
rubric---it is a structured AI-administered survey in which an assessor model produces six
component judgments, evidence excerpts, reasoning traces, and a confidence score for each of
nine cognitive dimensions. Because this instrument has grown sufficiently complex to carry its
own validation burden, the present study addresses the question prior to broad empirical
deployment.

\subsection{Background and Motivation}
\label{sec:motivation}

The MAAC framework \citep{DolehP2} introduced a process-oriented evaluation architecture that
decomposes AI cognitive performance into nine dimensions, each measured through a structured
six-component adjudication survey. The framework was theoretically grounded in cognitive-science
constructs and designed for empirical application across multiple AI models and task-complexity
conditions.

However, a theoretically grounded framework does not automatically produce a valid measurement
instrument. Before the MAAC instrument can support empirical claims about model-level cognitive
differences, the instrument itself must be validated. This requirement was motivated by concrete
diagnostic evidence. Early inspection of assessor output revealed that at least one dimension---
Tool Execution---was penalizing AI responses for not using tools even in conditions where tools
were intentionally absent. This is not an implementation bug; it is a construct-definition
problem that would invalidate downstream inferences if left unaddressed. Combined with wide
psychometric variability across the nine dimensions (preliminary Cronbach's $\alpha$ values
ranging from near-zero to above 0.90), these findings motivated a dedicated instrument-validation
study.

\subsection{The Central Question}
\label{sec:question}

This study is organized around a single methodological question:

\begin{quote}
\textit{Is the MAAC assessor survey fit for purpose?}
\end{quote}

This question is deliberately framed in the language of survey validation rather than
model-performance evaluation. Each MAAC dimension functions like a survey instrument: it presents
a structured set of six items to the assessor model, which responds with judgments, evidence, and
reasoning. The present study examines whether those instruments are well-constructed, internally
coherent, condition-aware, and ready for large-scale application---not whether any particular AI
model performs well on the dimensions.

\subsection{Research Questions}
\label{sec:rq}

Five research questions organize the study:

\begin{description}[noitemsep]
  \item[RQ1] \textbf{Assessment Architecture Validity.} Is the repeated six-component
        adjudication structure a coherent and defensible measurement architecture?
  \item[RQ2] \textbf{Dimension-Level Construct Coherence.} Do the six component assessments
        within each dimension behave like a coherent construct?
  \item[RQ3] \textbf{Condition Awareness.} Do dimensions that depend on available
        capabilities---especially Tool Execution---behave appropriately across no-tools and
        tools-enabled conditions?
  \item[RQ4] \textbf{Internal Agreement and Redundancy.} Do the six components provide
        meaningfully distinct structured evidence?
  \item[RQ5] \textbf{Measurement Readiness.} Is the revised MAAC assessor architecture ready
        for large-scale empirical application?
\end{description}

\subsection{Hypotheses}
\label{sec:hypotheses}

\begin{description}[noitemsep]
  \item[H1] \textbf{Six-Component Structural Coherence.} Each dimension's six component
        assessments will display acceptable internal consistency ($\alpha \geq 0.70$) and
        load on a single factor (CFI $\geq 0.90$).
  \item[H2] \textbf{Differential Dimension Quality.} Dimensions will vary meaningfully in
        psychometric quality, reflecting differences in construct complexity and observability.
  \item[H3] \textbf{Condition-Awareness Sensitivity.} Dimensions whose constructs depend on the
        active capability regime will produce score patterns that differ across conditions in
        construct-appropriate directions.
  \item[H4] \textbf{Adjudication Trace Diagnostic Value.} Component-level loading patterns will
        provide diagnostic information about construct ambiguity not visible from aggregate scores.
  \item[H5] \textbf{Revision Stability.} Targeted item revisions will not produce large
        unintended degradations ($d > 0.50$) in previously stable dimensions.
\end{description}

\subsection{Contributions}
\label{sec:contributions}

This study makes three contributions to AI evaluation methodology:

\begin{enumerate}[noitemsep]
  \item \textbf{The first psychometric validation of an AI-administered cognitive assessment
        instrument.} Rather than treating AI assessors as transparent scoring mechanisms, this
        study applies measurement-science rigor---internal consistency analysis, confirmatory
        factor analysis, schema validation, and targeted construct revision---to the AI
        adjudication layer itself.
  \item \textbf{A taxonomy of item-level failure modes in AI-administered instruments.} Five
        distinct failure categories are identified: ceiling/floor compression, construct drift,
        proxy-formula failure, condition confounding, and fundamental construct mismatch.
  \item \textbf{Empirical evidence that AI-administered instruments can be validated and improved
        using classical measurement-science principles.} The study demonstrates that CFA-driven
        diagnostic cycles yield actionable revision targets that reliably improve psychometric
        properties.
\end{enumerate}

% -------------------------------------------------------
% 2. BACKGROUND
% -------------------------------------------------------
\section{Background}
\label{sec:background}

\subsection{The Limits of Outcome-Based AI Evaluation}
\label{sec:limits}

The evaluation of large language models has developed primarily through outcome-based benchmarks.
Multi-task suites such as MMLU \citep{Hendrycks2021} and BIG-bench \citep{Srivastava2022} assess
accuracy across diverse domains. HELM \citep{Liang2022} added robustness, calibration, toxicity,
and efficiency. Despite this progress, these frameworks share a fundamental orientation: they
measure what models produce rather than how they produce it \citep{Mitchell2021, Rogers2020}.

Attempts to examine cognitive processes more directly have taken several forms: chain-of-thought
prompting \citep{Wei2022}, self-consistency methods \citep{Wang2022}, tree-of-thoughts
\citep{Yao2024}, and process supervision approaches \citep{Uesato2022}. Interpretability studies
examine internal representations and attention patterns \citep{Clark2019, DoshiVelez2017}.
Recent neuronal-level work has identified hallucination-associated neurons (H-Neurons) that
predict confabulation events, with fewer than 0.1\% of total model parameters driving the effect
\citep{Gao2025}; and knowledge neurons in pretrained transformers have been associated with
factual storage and retrieval \citep{Dai2022}. Each approach opens a window on process, but
externally generated reasoning traces may not faithfully reflect internal computation
\citep{Turpin2024, Saparov2023}, creating a gap between the aspiration of process-oriented
evaluation and the methodological infrastructure available to support it.

\subsection{The MAAC Framework}
\label{sec:maac}

The MAAC framework was introduced by \citet{DolehP2} as a theoretically grounded response to
the outcome dominance problem. Drawing on Marr's \citeyearpar{Marr1982} tri-level hypothesis,
Newell's \citeyearpar{Newell1990} unified theories of cognition, Anderson's
\citeyearpar{Anderson2004} ACT-R architecture, and Sweller's \citeyearpar{Sweller1988} cognitive
load theory, MAAC defines nine dimensions of AI cognitive processing. Table~\ref{tab:dimensions}
summarizes each dimension, its construct focus, and primary theoretical grounding.

\begin{table}[H]
\centering
\caption{MAAC Nine Dimensions: Construct Focus and Theoretical Grounding}
\label{tab:dimensions}
\small
\resizebox{\textwidth}{!}{%
\begin{tabular}{p{0.3cm}p{3cm}p{6cm}p{6cm}}
\toprule
\textbf{\#} & \textbf{Dimension} & \textbf{Construct Focus} & \textbf{Primary Grounding} \\
\midrule
1 & Cognitive Load           & Whether the response manages information burden and reasoning density appropriately for the task                           & \citet{Sweller1988} cognitive load; \citet{Baddeley1986} working memory \\
2 & Tool Execution           & Whether execution quality appropriately reflects the active tool and capability regime                                    & \citet{ClarkChalmers1998} extended cognition; \citet{Hutchins1995} \\
3 & Content Quality          & Whether the response is accurate, relevant, complete, and substantively useful                                           & Information quality frameworks; \citet{Mitchell2021} \\
4 & Memory Integration       & Whether the response appropriately uses or integrates memory and context resources                                       & \citet{Baddeley1986}; \citet{Anderson2004} ACT-R \\
5 & Complexity Handling      & Whether the response manages multi-step difficulty and complexity demands appropriately                                   & \citet{Campbell1988}; \citet{Newell1990} \\
6 & Hallucination Control    & Whether the response avoids unsupported fabrication and maintains epistemic discipline                                   & \citet{Turpin2024}; \citet{Dai2022}; \citet{Gao2025} \\
7 & Knowledge Transfer       & Whether knowledge is transferred appropriately across contexts and domains                                               & \citet{Anderson2004} ACT-R; \citet{Newell1990} \\
8 & Processing Efficiency    & Whether the response achieves useful output with appropriate process economy                                             & \citet{Marr1982}; \citet{Newell1990} \\
9 & Process-Outcome Alignment & Whether the response aligns process and output with the intended evaluative construct                                   & \citet{Messick1995}; \citet{Anderson2004} \\
\bottomrule
\end{tabular}}
\tablenote{Each dimension is operationalized through a six-component AI-administered adjudication survey. Dimension scores are computed as the mean of six component judgments scored on a 1--5 ordinal scale.}
\end{table}

Prior work established the theoretical foundations of MAAC and demonstrated content validity and
structural validity at the framework level \citep{DolehP2}. The nine-dimension framework builds
on a pre-validated scenario generation pipeline \citep{DolehP1} that established 4,238
complexity-controlled decision scenarios across four domains and three difficulty tiers.
What prior work did not do---and did not claim to do---was validate the adjudication instrument
through which those dimensions are measured. That task is the present study.

\subsection{The Need for Instrument Validation in AI-Based Adjudication}
\label{sec:need}

The use of AI models as evaluators of other AI models has grown substantially in recent years
\citep{Zheng2023}. However, the psychometric properties of AI-based adjudication have received
limited systematic study \citep{Sottovia2024, Hada2023}. Building on Messick's
\citeyearpar{Messick1995} unified validity theory and the COSMIN framework
\citep{Mokkink2010, Terwee2024}, instrument validation typically requires evidence of:
content validity; internal structure validity; reliability; condition-sensitivity; and
discriminant validity. These requirements translate directly to testable claims about the MAAC
adjudication instrument.

\subsection{Evidence Motivating the Present Study}
\label{sec:motivation_evidence}

Three classes of evidence motivated a dedicated instrument-validation study.

\textbf{First, substantial psychometric heterogeneity.} Preliminary Cronbach's $\alpha$ estimates
ranged from approximately 0.008 (Complexity Handling) and 0.026 (Hallucination Control) to 0.758
(Tool Execution) and 0.903 (Knowledge Transfer). Values below 0.60 are not defensible for scale
use \citep{Nunnally1994, DeVellis2017}; values near zero indicate that the six component items
within those dimensions are effectively unrelated.

\textbf{Second, Tool Execution showed a clearly condition-inappropriate scoring pattern.}
In the no-tools baseline condition---where the experimental design intentionally withheld
external tools---Tool Execution scores collapsed to substantially lower values than in
conditions where tools were available. Because tool non-use was required by the experimental
condition, these score differences cannot be attributed to execution quality. They reveal a
construct-definition problem \citep{DolehP1, DolehTR2026a}.

\textbf{Third, qualitative review of the six-component adjudication traces showed that the
assessor's reasoning and evidence selection were materially useful for diagnosing construct
ambiguity}, suggesting that the trace structure provides interpretable diagnostic information
even when top-line scores are problematic.

Table~\ref{tab:preliminary} summarizes the preliminary internal-consistency diagnostics across
all nine dimensions.

\begin{table}[H]
\centering
\caption{Preliminary Internal Consistency Diagnostics by MAAC Dimension ($n = 599$, Pre-Revision)}
\label{tab:preliminary}
\small
\resizebox{\textwidth}{!}{%
\begin{tabular}{lccc}
\toprule
\textbf{Dimension} & \textbf{Cronbach's $\alpha$} & \textbf{Mean Item-Total $r$} & \textbf{Preliminary Status} \\
\midrule
Knowledge Transfer      & 0.903 & 0.759 & Strong --- meets threshold \\
Memory Integration      & 0.763 & 0.496 & Moderate --- acceptable, monitor \\
Process-Outcome Alignment & 0.763 & 0.579 & Moderate --- acceptable, monitor \\
Tool Execution          & 0.758 & 0.534 & Moderate --- construct-definition concern \\
Processing Efficiency   & 0.702 & 0.447 & Moderate --- borderline, monitor \\
Content Quality         & 0.544 & 0.316 & Weak --- below threshold \\
Cognitive Load          & 0.291 & 0.207 & Very weak --- below threshold \\
Hallucination Control   & 0.026 & 0.071 & Near-zero --- items unrelated \\
Complexity Handling     & 0.008 & 0.061 & Near-zero --- items unrelated \\
\bottomrule
\end{tabular}}
\tablenote{Diagnostics from the pre-revision, pre-schema-enforcement batch ($n = 599$). Values informed Stage~A construct audit priorities and are not treated as formal validation outcomes. Threshold: $\alpha \geq 0.70$ \citep{Nunnally1994}. Tool Execution meets the consistency threshold but independently requires construct revision due to condition-awareness failure.}
\end{table}

\subsection{Measurement Validation Theory}
\label{sec:theory}

The validation program draws on two established frameworks. \textbf{Messick's
\citeyearpar{Messick1995} unified validity theory} treats validity as a property of the
inferences drawn from test scores. Validation requires accumulating evidence from multiple
sources to support the inferential claims associated with a measurement instrument. Applied here:
each MAAC dimension makes an inferential claim (e.g., ``this score reflects the degree to which
the response manages cognitive load appropriately''), and validating that claim requires
demonstrating that the instrument measures what it claims, not something adjacent.

\textbf{The COSMIN framework} \citep{Mokkink2010, Terwee2024} operationalizes measurement
validation for instruments used in empirical research, providing explicit criteria for content
validity, structural validity, internal consistency, and reliability
\citep{Prinsen2023, Murphy2024}. Building on classical psychometric theory
\citep{Cronbach1951, McDonald1999}, we treat the six component assessments within each MAAC
dimension as a multi-item scale measuring a latent construct. Internal consistency and factor
analysis \citep{Fabrigar1999, Brown2015} provide the primary empirical evidence.

% -------------------------------------------------------
% 3. METHODS
% -------------------------------------------------------
\section{Methods}
\label{sec:methods}

\subsection{Study Design}
\label{sec:design}

The present study is an instrument-validation study targeting the MAAC adjudication instrument.
The design combines conceptual validation (construct definition review and item-level audit) with
empirical validation (psychometric testing of the six-component architecture in a controlled
validation dataset). It proceeds through five sequential stages: conceptual construct audit
(Stage~A), targeted assessor revision (Stage~B), validation data collection (Stage~C),
statistical validation (Stage~D), and readiness decision (Stage~E). This design follows
established procedures for instrument development and validation
\citep{DeVellis2017, Furr2018, Mokkink2010}.

The scope is intentionally narrow. The present study does not pursue comprehensive substantive
claims about target model performance, nor does it generate the full application dataset needed
for multi-model comparison. Its scope is limited to determining whether the measurement
instrument is fit for the purposes for which it will be used in subsequent studies.

\subsection{The MAAC Assessor Architecture}
\label{sec:architecture}

The MAAC adjudication instrument is administered by a large language model (the ``assessor'')
that reads each target AI response and applies a structured scoring prompt for each of the nine
MAAC dimensions. For each dimension, the assessor produces six component judgments scored on a
1--5 ordinal scale, component evidence excerpts, component reasoning, a dimension score
(the arithmetic mean of the six components), and a dimension confidence score.

Under this architecture, the six component judgments function as items in a multi-item scale.
The dimension score is:
$$D_i = \frac{1}{6}\sum_{j=1}^{6} q_{ij}$$
where $D_i$ is the score for dimension~$i$ and $q_{ij}$ is the score for component~$j$ of that
dimension. The assessor model used throughout is DeepSeek-Chat V3.

\begin{figure}[H]
  \centering
  \includegraphics[width=\textwidth]{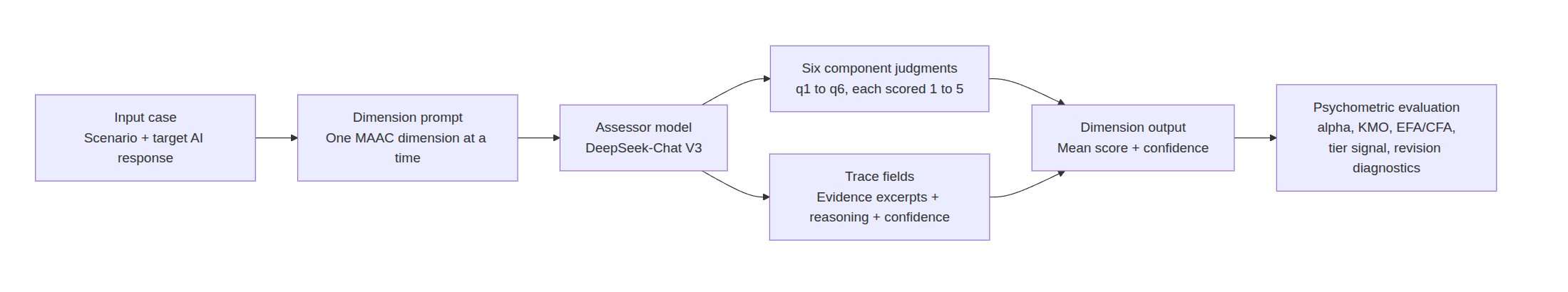}
  \caption{MAAC adjudication architecture. Input cases (scenario + target AI response) are
    processed dimension-by-dimension through the assessor model (DeepSeek-Chat V3), which
    produces six component judgments ($q_1$--$q_6$, each scored 1--5) alongside trace fields
    (evidence excerpts, reasoning, confidence). The dimension output---mean score plus
    confidence---serves as input to psychometric evaluation including $\alpha$, KMO, EFA/CFA,
    tier signal, and revision diagnostics.}
  \label{fig:architecture}
\end{figure}

Figure~\ref{fig:architecture} illustrates the full adjudication pipeline from input case to
psychometric evaluation.

Table~\ref{tab:facets} provides the full component specification, listing the labeled facet
assessed by each of the six component questions across all nine MAAC dimensions.

\begin{table}[H]
\centering
\caption{Six-Component Facet Specification by MAAC Dimension}
\label{tab:facets}
\small
\resizebox{\textwidth}{!}{%
\begin{tabular}{p{2.5cm}p{2.8cm}p{2.8cm}p{2.8cm}p{2.8cm}p{2.8cm}p{2.8cm}p{3.5cm}}
\toprule
\textbf{Dimension} & \textbf{q1} & \textbf{q2} & \textbf{q3} & \textbf{q4} & \textbf{q5} & \textbf{q6} & \textbf{Key References} \\
\midrule
Cognitive Load        & Concepts per segment (Miller's 7±2) & Primary task allocation ($>$60\% effort) & Information density for task complexity & Cognitive overload indicators & Processing efficiency by complexity tier & Resource optimization vs.\ tool regime & \citet{Sweller1988}; \citet{Baddeley1986}; \citet{Miller1956} \\
Tool Execution        & Tool appropriateness for task requirements & Execution quality and parameter configuration & Tool output integration into response & Execution efficiency (no redundant calls) & Success-criteria achievement under active capability regime & Capability-appropriate proportionate tool use & \citet{ClarkChalmers1998}; \citet{Hutchins1995}; \citet{Schick2024}; \citet{Nakano2021} \\
Content Quality       & Factual accuracy rate & Content relevance to task & Clarity and structural quality & Completeness for task scope & Success criteria achievement & Domain expertise demonstrated & \citet{Bevilacqua2023}; \citet{Ghildyal2025} \\
Memory Integration    & Factual claim connection to context & Semantic continuity across response & Explicitness of knowledge connections & Contextual appropriateness of retrieved information & Memory utilization efficiency & Storage-retrieval balance for coherence & \citet{Baddeley1986}; \citet{Anderson2004}; \citet{AMem2025}; \citet{MemOS2025} \\
Complexity Handling   & Multi-component problem decomposition & Solution step sequencing quality & Cognitive complexity appropriateness & Edge case and complication coverage & Alignment with complexity tier demands & Cognitive resource allocation efficiency & \citet{Campbell1988}; \citet{Newell1990}; \citet{Joksimovic2023}; \citet{Mitrai2024} \\
Hallucination Control & Uncertainty marker appropriateness & Claim qualification with confidence levels & Internal consistency (absence of contradictions) & Distinction between speculation and fact & Processing calibration for verification thoroughness & Evidential calibration under epistemic risk & \citet{Turpin2024}; \citet{Saparov2023}; \citet{Ji2023}; \citet{Dai2022}; \citet{Gao2025} \\
Knowledge Transfer    & Cross-domain pattern transfer validity & Abstraction and generalization quality & Specialized knowledge adaptation for context & Novel inter-concept connection quality & Transfer opportunity coverage and adaptive application & Domain-tailored adaptive insight generation & \citet{Anderson2004}; \citet{Newell1990}; \citet{Savelka2021} \\
Processing Efficiency & Information density without clarity loss & Reasoning step economy & Resource allocation by component importance & Redundancy control while maintaining coherence & Processing speed relative to output quality & Cognitive and memory operational efficiency & \citet{Marr1982}; \citet{Sweller1988}; \citet{Newell1990} \\
Process-Outcome Alignment & Internal consistency throughout response & Logical validity of conclusions from premises & Theoretical coherence for domain & Definitional consistency across response & Construct alignment with intended evaluation purpose & Process-outcome alignment for domain-specific constructs & \citet{Messick1995}; \citet{Anderson2004}; \citet{Campbell1959} \\
\bottomrule
\end{tabular}}
\tablenote{Component labels reflect the best accepted revision state at the end of the present study, including grounded revisions for Tool Execution and Knowledge Transfer and the \texttt{evidential\_calibration} replacement for Hallucination Control Q6. Each component is scored on a 1--5 ordinal scale by the AI assessor.}
\end{table}

\subsection{Validation Dataset Design}
\label{sec:dataset}

The validation dataset is drawn from the validated scenario pool established in prior work
\citep{DolehP1}, containing over 4,000 complexity-validated decision scenarios distributed
across four domains (analytical, planning, communication, problem-solving) and three complexity
tiers (low, medium, high). The stratified validation design included all four domains, all three
complexity tiers, and at least one no-tools condition (Condition~A) providing the primary dataset
for psychometric analysis.

\begin{table}[H]
\centering
\caption{Original Validation Sample Design: Stratified $4 \times 3 \times 2$ Matrix (Per Condition)}
\label{tab:design}
\small
\resizebox{\textwidth}{!}{%
\begin{tabular}{lccccc}
\toprule
\textbf{Complexity Tier} & \textbf{Analytical} & \textbf{Planning} & \textbf{Communication} & \textbf{Problem-Solving} & \textbf{Row Total} \\
\midrule
Low    & 25 & 25 & 25 & 25 & 100 \\
Medium & 25 & 25 & 25 & 25 & 100 \\
High   & 25 & 25 & 25 & 25 & 100 \\
\midrule
\textbf{Column Total} & \textbf{75} & \textbf{75} & \textbf{75} & \textbf{75} & \textbf{300} \\
\bottomrule
\end{tabular}}
\tablenote{Cell counts are per condition in the original validation design. The completed evidence base ultimately exceeded this initial checkpoint: full baseline run ($n = 2{,}196$) and a final HC~Q6 confirmation cohort ($n = 1{,}774$). The matched tools-enabled comparison remains future work.}
\end{table}

\textit{Note on unit of analysis.} Each completed assessed response contributes one six-item
dimension record for factor-analytic purposes. EFA/CFA does not require 300 different AI models;
it requires a sufficiently large sample of adjudicated survey instances exhibiting meaningful
variation in the constructs being measured. The psychometric claims should therefore be framed as
validation of the instrument under an AI adjudication regime, not as population psychology claims
about AI models as cognitive agents.

\subsection{Statistical Validation Plan (Stage D)}
\label{sec:stats}

Stage~D proceeds through seven sequential steps (Table~\ref{tab:validation_plan}).

\begin{table}[H]
\centering
\caption{Stage~D Statistical Validation Plan: Analysis Steps, Methods, and Success Criteria}
\label{tab:validation_plan}
\small
\resizebox{\textwidth}{!}{%
\begin{tabular}{p{0.5cm}p{3cm}p{3cm}p{5cm}p{3cm}}
\toprule
\textbf{Step} & \textbf{Analysis} & \textbf{Method} & \textbf{Success Criterion} & \textbf{Addresses} \\
\midrule
1 & Descriptive inspection   & Score distributions, frequency tables & No floor/ceiling effect ($< 15\%$ at extreme values) & H2, H3 \\
2 & Internal consistency     & Cronbach's $\alpha$, McDonald's $\omega$, mean inter-item $r$, corrected item-total $r$ & $\alpha \geq 0.70$, $\omega \geq 0.70$ & H1, H2 \\
3 & Redundancy analysis      & Pairwise Pearson $r$ among six components & No component pair $r > 0.85$ & H4, RQ4 \\
4 & Exploratory factor analysis & ML extraction, Promax rotation, parallel analysis & One-factor; all loadings $\geq 0.40$ & H1, RQ1 \\
5 & Confirmatory factor analysis & Single-factor CFA via semopy & CFI $\geq 0.90$; TLI $\geq 0.90$; RMSEA $\leq 0.08$; SRMR $\leq 0.08$ & H1, RQ2 \\
6 & Condition-awareness analysis & Mann-Whitney $U$ / $t$-test; ANOVA by tier & Theoretically coherent condition differences & H3, RQ3 \\
7 & Readiness classification & Multi-criterion synthesis of Steps 1--6 & $\geq 6$ of 9 dimensions Ready or Conditionally Ready & RQ5, H1--H3 \\
\bottomrule
\end{tabular}}
\tablenote{Steps proceed sequentially. In the executed study, Steps~1--5 and~7 were completed on the baseline validation and revision cohorts, while Step~6 was addressed indirectly through targeted revision evidence. Significance threshold $\alpha = .05$ throughout; effect sizes reported for all inferential tests.}
\end{table}

\textbf{Readiness classification criteria} are defined in Table~\ref{tab:readiness_criteria}.

\begin{table}[H]
\centering
\caption{Dimension Readiness Classification Criteria}
\label{tab:readiness_criteria}
\small
\resizebox{\textwidth}{!}{%
\begin{tabular}{p{2.5cm}p{3.5cm}p{3.5cm}p{3.5cm}p{3.5cm}}
\toprule
\textbf{Classification} & \textbf{Internal Consistency} & \textbf{Factor Structure} & \textbf{Condition Behavior} & \textbf{Disposition} \\
\midrule
\textbf{Ready}              & $\alpha \geq 0.70$ and $\omega \geq 0.70$ & CFI $\geq 0.90$, RMSEA $\leq 0.08$, SRMR $\leq 0.08$ & Theoretically coherent across all conditions & Proceed to empirical application without restriction \\
\textbf{Conditionally Ready} & $\alpha$ 0.50--0.69 or one threshold met near the other & CFI $\geq 0.85$ with interpretable loadings; or EFA-only support & Largely coherent; minor anomalies that do not invalidate the construct & Proceed with documented caveats; flag for monitoring \\
\textbf{Not Ready}          & $\alpha < 0.50$ after revision, or item-total $r < 0.20$ for majority & Factor structure indefensible or CFA convergence failure & Condition-inappropriate collapse or reversal persisting after revision & Requires further reconceptualization \\
\bottomrule
\end{tabular}}
\end{table}

\subsection{Iterative Revision Methodology}
\label{sec:iterative}

The initial validation design assumed a single revision pass would suffice. Round~1 results
revealed the gap was too large to close in one iteration. The study therefore adopted an
iterative methodology: after each validation run ($n \approx 2{,}190$), CFA was conducted,
items loading below 0.40 were flagged, and each was diagnosed through score distribution
inspection, corrected item-total correlations, inter-item correlation matrices, and qualitative
review of assessor reasoning traces. Diagnoses were classified into five failure types:
(a)~ceiling/floor compression, (b)~construct drift, (c)~proxy-formula failure,
(d)~condition confounding, and (e)~fundamental construct mismatch. Targeted revisions were
implemented and validated on fresh cohorts. Eight instrument states were evaluated across the
full program: the v4.0 baseline, six major revision rounds through v4.6, and a final v4.7
confirmation targeting HC~Q6.

% -------------------------------------------------------
% 4. RESULTS
% -------------------------------------------------------
\section{Results}
\label{sec:results}

\subsection{Validation Workflow and Dataset Progression}
\label{sec:workflow}

\begin{figure}[H]
  \centering
  \includegraphics[width=\textwidth]{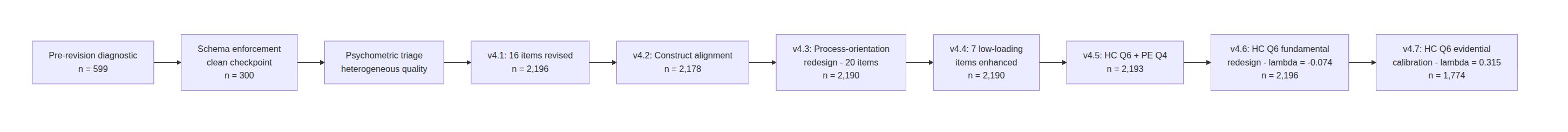}
  \caption{Validation workflow from Round~1 baseline through iterative revision to current
    accepted state. Each box represents a major instrument state or revision cycle with its
    validation cohort size. The program progressed from a pre-revision diagnostic
    ($n = 599$) through schema enforcement, psychometric triage, and seven revision cycles
    (v4.1--v4.7), culminating in the final HC~Q6 \texttt{evidential\_calibration} redesign
    ($\lambda = 0.315$, $n = 1{,}774$).}
  \label{fig:workflow}
\end{figure}

The validation proceeded as a multi-round iterative validation-and-revision program rather than
a single static validation pass. Figure~\ref{fig:workflow} shows the progression from the
pre-revision diagnostic through seven revision cycles to the final v4.7 instrument state.

\subsection{Schema Integrity}
\label{sec:schema}

The first decisive result was structural. In the pre-revision diagnostic dataset ($n = 599$),
the six-item survey structure was not consistently enforced: malformed dimension traces,
extra component keys, and missing-item cases were all present, invalidating any attempt to
treat that batch as the final validation dataset.

After schema enforcement, a fresh balanced validation batch ($n = 300$) showed perfect
structural integrity: 0 malformed traces, 0 extra-key violations, 0 missing-item violations.
All later psychometric claims depend on a stable survey contract. Once the six-item structure
was enforced cleanly, the study could move from ``is the survey even being emitted correctly?''
to ``is the survey substantively measuring coherent constructs?''

\subsection{Round 1 Psychometric Baseline}
\label{sec:baseline}

The clean post-schema checkpoint showed strong differential dimension quality.
Table~\ref{tab:round1} consolidates the Round~1 readiness evidence.

\begin{table}[H]
\centering
\caption{Round~1 Readiness Synthesis and Revision Priorities ($n = 594$ Full Trace Dataset)}
\label{tab:round1}
\small
\resizebox{\textwidth}{!}{%
\begin{tabular}{p{2.8cm}ccccp{4cm}p{2.5cm}}
\toprule
\textbf{Dimension} & \textbf{$\alpha$ ($n=300$)} & \textbf{$\alpha$ ($n=594$)} & \textbf{KMO} & \textbf{Key Diagnostic Finding} & \textbf{Revision Status} & \textbf{Readiness} \\
\midrule
Knowledge Transfer      & 0.903 & \textbf{0.945} & 0.912 (marvelous) & All 6 item-total $r > 0.74$; textbook unidimensional structure  & Grounded Q5/Q6 accepted     & \textbf{Ready} \\
Process-Outcome Alignment & 0.763 & \textbf{0.722} & 0.750 (middling)  & Q3/Q5/Q6 strong subcore; Q2/Q4 positive but weaker              & No targeted revision        & \textbf{Ready} \\
Processing Efficiency   & 0.702 & 0.621           & 0.607 (mediocre)  & Q4 item-total $r \approx 0.000$; 5-item subcore coherent        & No targeted revision        & \textbf{Conditionally Ready} \\
Memory Integration      & 0.763 & 0.610           & 0.655 (mediocre)  & Q2 item-total $r = -0.013$; 5-item subcore coherent             & No targeted revision        & \textbf{Conditionally Ready} \\
Complexity Handling     & 0.008$^\dagger$ & 0.538 & 0.602 (mediocre)  & All 6 items positively loading; shallow shared variance         & Schema-fix artifact resolved & \textbf{Conditionally Ready} \\
Hallucination Control   & 0.026$^\dagger$ & 0.527 & 0.686 (mediocre)  & Q1--Q4 subcore viable; Q5/Q6 divergent or orthogonal            & Schema-fix artifact resolved & \textbf{Conditionally Ready} \\
Content Quality         & 0.544 & 0.526           & 0.604 (mediocre)  & Q1/Q3 inert (item-total $r < 0.04$); Q2/Q4/Q5 subcore viable  & No targeted revision        & \textbf{Conditionally Ready} \\
Tool Execution          & 0.758$^\ddagger$ & 0.107$^\ddagger$ & 0.500 (miserable) & Post-revision ceiling artifact; grounded revision accepted & Grounded Q6 accepted       & \textbf{Conditionally Ready} \\
Cognitive Load          & 0.291 & \textbf{$-0.152$} & 0.559 (miserable) & Negative $\alpha$; no coherent latent dimension; fundamental redesign required & No targeted redesign & \textbf{Not Ready} \\
\bottomrule
\end{tabular}}
\tablenote{$\alpha$ values for $n=300$ are from the post-schema-enforcement checkpoint. $\alpha$ values for $n=594$ are from the full baseline validation run. KMO $> 0.60$ indicates adequate factorability. $^\dagger$Near-zero $\alpha$ for Complexity Handling and Hallucination Control at $n=300$ were schema-corruption artifacts; post-schema values at $n=594$ reflect the true construct state. $^\ddagger$Tool Execution $\alpha = 0.758$ at $n=300$ reflected post-schema checkpoint data; $\alpha = 0.107$ at $n=594$ is a post-revision ceiling artifact. The grounded-revision cohort (M~$= 4.772$, SD~$= 0.420$, $n = 600$) is the primary evidence for TE readiness.}
\end{table}

The tier-differentiation signal at $n = 594$ was large: overall MAAC $F(2, 591) = 114.67$,
$p < .001$, $\eta^{2} = 0.280$. This confirms that the assessment method is responsive to
structured task-complexity variation.

\subsection{Revision Case Study 1: Tool Execution}
\label{sec:te}

Tool Execution was the clearest initial construct failure. The pre-revision instrument penalized
no-tools baseline responses for not using tools even when tool non-use was mandated by the
experimental condition. The accepted revision had two core features: (1)~explicit
baseline/enhanced condition logic, so baseline non-use was not treated as execution failure;
(2)~a grounded Q6 formulation based on opportunity structure, relevant tool use, redundancy
control, and missed tool opportunities.

\begin{table}[H]
\centering
\caption{Accepted Revision Outcomes for the Two Targeted Dimensions}
\label{tab:revision}
\small
\resizebox{\textwidth}{!}{%
\begin{tabular}{p{2.5cm}p{3cm}p{3cm}p{4cm}p{4cm}p{2cm}}
\toprule
\textbf{Dimension} & \textbf{Revision Focus} & \textbf{Comparison Basis} & \textbf{Pre-Revision Result} & \textbf{Post-Revision Result} & \textbf{Decision} \\
\midrule
Tool Execution    & Replace condition-confounded back-half logic with grounded capability-aware Q6 & Earlier post-Q5 checkpoint vs.\ full grounded cohort & M~$= 4.238$, SD~$= 0.459$, Q6 low-tail heavy ($n = 202$) & M~$= 4.772$, SD~$= 0.420$, Q6 $= 5$ modal ($n = 600$) & Keep grounded TE as current best version \\
Knowledge Transfer & Replace proxy-heavy Q5/Q6 with grounded transfer-opportunity and adaptive-insight metrics & Pre-grounded KT vs.\ first grounded KT validation & M~$= 3.733$, SD~$= 0.565$, Q5/Q6 compressed ($n = 600$) & M~$= 3.936$, SD~$= 0.795$, Q5 materially decompressed ($n = 597$) & Keep grounded KT as current best version \\
\bottomrule
\end{tabular}}
\end{table}

\subsection{Iterative Revision Program: v4.0 Through v4.6}
\label{sec:iterations}

Table~\ref{tab:versions} summarizes the scope, rationale, and key outcomes of each instrument
version across the full iterative program.

\begin{table}[H]
\centering
\caption{Instrument Version Progression: v4.0 Through v4.6}
\label{tab:versions}
\small
\resizebox{\textwidth}{!}{%
\begin{tabular}{p{1cm}p{2cm}p{0.8cm}p{5cm}p{1.5cm}p{5cm}}
\toprule
\textbf{Version} & \textbf{Date} & \textbf{Items} & \textbf{Key Changes} & \textbf{$n$} & \textbf{Primary Outcome} \\
\midrule
v4.0 & Mar 11, 2026 & --- & Baseline instrument & 2,196 & Round~1 diagnostic: $\alpha$ range $-0.152$ to $0.945$; 1/9 CFI $\geq 0.95$ \\
v4.1 & Mar 11, 2026 & 16  & MI Q1--Q4 full overhaul; CL Q2--Q6; CQ Q1/Q3 & 2,196 & Mixed: MI improved; CL remained weak \\
v4.2 & Mar 12, 2026 & 12  & Construct alignment audit: CQ redefined (vocabulary, discourse, register); PE as ``response economy'' & 2,178 & CQ construct aligned to framework definitions \\
v4.3 & Mar 12--13, 2026 & 20 & Process-orientation redesign: non-monotonic scoring, tier calibration; KT full redesign; TE full rewrite; POA Q1--Q5 & 2,190 & 48/48 non-CQ items $= 100\%$ process-oriented; KT $\alpha$ 0.945$\to$0.920 \\
v4.4 & Mar 13, 2026 & 7   & Targeted low-loading items: POA Q3, MI Q5/Q6, CH Q6, CL Q4, TE Q6, HC Q6 & 2,190 & 6/7 targets fixed (all $\geq 0.40$); PE first CFA convergence (CFI $= 0.975$); HC Q6 regressed \\
v4.5 & Mar 14, 2026 & 2   & HC Q6 binary simplification; PE Q4 gradient scoring & 2,193 & PE Q4 fixed ($0.393 \to 0.458$); HC Q6 worsened ($0.276 \to 0.169$) \\
v4.6 & Mar 14, 2026 & 1   & HC Q6 fundamental redesign: \texttt{source\_calibration} $\to$ \texttt{analytical\_confidence\_calibration} & 2,196 & HC Q6 $\lambda = -0.074$ (worse); all other dimensions stable ($|d| < 0.035$) \\
\bottomrule
\end{tabular}}
\end{table}

\textbf{Reliability progression.} Table~\ref{tab:alpha_progression} tracks Cronbach's $\alpha$
across major revision versions.

\begin{table}[H]
\centering
\caption{Internal Consistency Progression: Cronbach's $\alpha$ by Dimension Across Instrument Versions}
\label{tab:alpha_progression}
\small
\resizebox{\textwidth}{!}{%
\begin{tabular}{lccccccc}
\toprule
\textbf{Dimension} & \textbf{v4.0} & \textbf{v4.1} & \textbf{v4.2} & \textbf{v4.3} & \textbf{v4.4} & \textbf{v4.5} & \textbf{$\Delta$ (v4.0$\to$v4.5)} \\
\midrule
CL  & $-0.152$ & 0.764 & 0.729 & 0.764 & 0.795 & 0.765 & \textbf{+0.917} \\
TE  &  0.107   & 0.789 & 0.767 & 0.789 & 0.808 & 0.794 & \textbf{+0.687} \\
CQ  &  0.526   & 0.821 & 0.824 & 0.821 & 0.801 & 0.825 &         +0.299  \\
MI  &  0.610   & 0.751 & 0.741 & 0.751 & 0.822 & 0.841 &         +0.231  \\
CH  &  0.538   & 0.740 & 0.726 & 0.740 & 0.762 & 0.764 &         +0.226  \\
HC  &  0.527   & 0.896 & 0.889 & 0.896 & 0.891 & 0.881 &         +0.354  \\
KT  &  0.945   & ---   & 0.945 & 0.923 & 0.920 & 0.924 &         $-0.021$ \\
PE  &  0.621   & 0.800 & 0.802 & 0.800 & 0.834 & 0.848 &         +0.227  \\
POA & ---       & 0.780 & 0.758 & 0.780 & 0.805 & 0.784 &         ---      \\
\midrule
\textbf{Mean} & --- & --- & --- & 0.807 & \textbf{0.829} & \textbf{0.825} & --- \\
\bottomrule
\end{tabular}}
\tablenote{v4.0 values from the Round~1 baseline ($n = 594$ component trace dataset). KT declined slightly due to deliberate shift from output-quality ratios to cognitive-process metrics in v4.3. POA v4.0 baseline not reported due to concurrent schema and construct changes. All v4.5 dimensions meet $\alpha \geq 0.70$.}
\end{table}

\subsection{Historical v4.5 CFA Checkpoint ($n = 2{,}193$)}
\label{sec:v45}

Table~\ref{tab:cfa_v45} presents the comprehensive CFA results for the v4.5 checkpoint.

\begin{table}[H]
\centering
\caption{Historical v4.5 Confirmatory Factor Analysis Checkpoint ($n = 2{,}193$)}
\label{tab:cfa_v45}
\small
\resizebox{\textwidth}{!}{%
\begin{tabular}{lcccccccccccc}
\toprule
\textbf{Dim.} & \textbf{$\alpha$} & \textbf{$\omega$} & \textbf{CFI} & \textbf{Fit} & \textbf{RMSEA} & \textbf{Q1 $\lambda$} & \textbf{Q2 $\lambda$} & \textbf{Q3 $\lambda$} & \textbf{Q4 $\lambda$} & \textbf{Q5 $\lambda$} & \textbf{Q6 $\lambda$} \\
\midrule
CL  & 0.765 & 0.758 & 0.878 & Poor    & 0.153 & 0.830 & 0.659 & 0.665 & 0.403 & 0.455 & 0.658 \\
TE  & 0.794 & 0.825 & 0.929 & Accept. & 0.121 & 0.759 & 0.758 & 0.627 & 0.613 & 0.700 & \textbf{0.391} \\
CQ  & 0.825 & 0.753 & 0.823 & Poor    & 0.228 & 0.759 & 0.563 & 0.654 & 0.722 & 0.509 & 0.877 \\
MI  & 0.841 & 0.832 & 0.912 & Accept. & 0.162 & 0.546 & 0.756 & 0.753 & 0.875 & 0.731 & 0.551 \\
CH  & 0.764 & 0.807 & 0.933 & Accept. & 0.106 & 0.587 & 0.566 & 0.827 & 0.582 & 0.604 & 0.508 \\
HC  & 0.881 & 0.866 & \textbf{0.971} & \textbf{Good} & 0.125 & 0.958 & 0.954 & 0.844 & 0.718 & 0.825 & \textbf{0.169} \\
KT  & 0.924 & 0.901 & \textbf{0.955} & \textbf{Good} & 0.162 & 0.916 & 0.675 & 0.808 & 0.778 & 0.895 & 0.933 \\
PE  & 0.848 & 0.848 & 0.932 & Accept. & 0.141 & 0.729 & 0.860 & 0.756 & 0.458 & 0.748 & 0.658 \\
POA & 0.784 & 0.801 & 0.934 & Accept. & 0.134 & 0.545 & 0.954 & 0.600 & 0.799 & \textbf{0.386} & 0.678 \\
\bottomrule
\end{tabular}}
\tablenote{\textbf{Bold} loadings indicate items below the 0.40 threshold. CFI Fit: Good ($\geq 0.95$), Acceptable ($\geq 0.90$), Poor ($< 0.90$). All RMSEA values exceed 0.08, consistent with known sensitivity of this index to low-df ($df = 9$) models at large sample sizes \citep{Kenny2015}. SRMR values are reasonable (0.03--0.08 range) across all dimensions.}
\end{table}

\subsection{HC Q6: Fundamental Redesign and Partial Recovery (v4.6--v4.7)}
\label{sec:hcq6}

Hallucination Control~Q6 remained the only critical blocker after five revision cycles.
Its loading trajectory through v4.6 was unique: v4.3 $= -0.397$, v4.4 $= 0.276$,
v4.5 $= 0.169$, v4.6 $= -0.074$.

Deep statistical analysis of the v4.5 data ($n = 2{,}193$) confirmed a fundamental construct
mismatch. Q6's corrected item-total correlation was 0.198 (vs.\ 0.492--0.812 for Q1--Q5). In
the complex tier, Q6 correlated \textit{negatively} with Q1 ($r = -0.218$, $p < .002$),
indicating that high source-attribution scores were associated with \textit{lower} epistemic
calibration performance. Source attribution represents a categorically different cognitive
behavior from epistemic calibration.

The v4.7 redesign replaced {\small\texttt{analytical\_confidence\_calibration}} with
{\small\texttt{evidential\_calibration}}, returning to the best prior v4.3 logic while reframing the
item as a behavioral complement to Q1--Q5 verbal epistemic calibration. HC~Q6 loaded at
$\lambda = 0.315$---improving over v4.5 (0.169), v4.6 ($-0.074$), and the prior best v4.3
result (0.308)---but remaining below the 0.40 target threshold.

\textit{Revision stability (H5).} The v4.7 versus v4.6 stability export showed statistically
detectable but substantively trivial cross-version movement: overall MAAC $d = -0.084$,
with no individual dimension exceeding $d = 0.118$.

\subsection{Tier Differentiation Signal}
\label{sec:tier}

The full $n = 594$ component trace dataset examined whether the MAAC instrument differentiates
across complexity tiers. Table~\ref{tab:tier} reports one-way ANOVA results by dimension.

\begin{table}[H]
\centering
\caption{Tier Differentiation Signal: One-Way ANOVA Results by Dimension ($n = 594$)}
\label{tab:tier}
\small
\resizebox{\textwidth}{!}{%
\begin{tabular}{lccccccp{4.5cm}}
\toprule
\textbf{Dimension} & \textbf{$F$} & \textbf{$p$} & \textbf{$\eta^{2}$} & \textbf{Simple M} & \textbf{Moderate M} & \textbf{Complex M} & \textbf{Signal} \\
\midrule
Knowledge Transfer & 205.90 & $<.001$ & \textbf{0.411} & 3.140 & 4.035 & 4.354 & $\uparrow$ monotonic --- largest differentiator \\
Cognitive Load     &  80.61 & $<.001$ & 0.214          & 4.440 & 4.347 & 4.877 & Non-monotonic (moderate dips) \\
Complexity Handling &  63.49 & $<.001$ & 0.177          & 3.955 & 4.090 & 4.410 & $\uparrow$ monotonic \\
Processing Efficiency & 52.66 & $<.001$ & 0.151        & 3.805 & 4.080 & 4.246 & $\uparrow$ monotonic \\
Content Quality    &  32.17 & $<.001$ & 0.098          & 4.515 & 4.784 & 4.877 & $\uparrow$ monotonic \\
Memory Integration &   7.25 & .001   & 0.024           & 4.015 & 4.156 & 4.164 & $\uparrow$ (plateau at moderate) \\
Hallucination Control & 5.35 & .005  & 0.018           & 4.035 & 4.085 & 4.174 & $\uparrow$ monotonic (small) \\
Tool Execution     &   1.57 & .209 (ns) & 0.005        & 4.795 & 4.729 & 4.728 & Near-ceiling --- no tier signal \\
Process-Outcome Alignment & 0.19 & .827 (ns) & 0.001  & 4.420 & 4.442 & 4.451 & Near-ceiling --- no tier signal \\
\midrule
\textbf{Overall MAAC} & \textbf{114.67} & \textbf{$<.001$} & \textbf{0.280} & \textbf{4.124} & \textbf{4.303} & \textbf{4.475} & \textbf{$\uparrow$ monotonic} \\
\bottomrule
\end{tabular}}
\tablenote{Groups: simple $n = 200$, moderate $n = 199$, complex $n = 195$. $\eta^{2}$ benchmarks: $< 0.06$ small, 0.06--0.14 moderate, $> 0.14$ large \citep{Cohen1988}. Analysis is exploratory; used to assess method-level sensitivity, not to make dimension-level validity claims.}
\end{table}

\subsection{Pre/Post Instrument Comparison (H5)}
\label{sec:prepost}

A full-scale pre/post comparison between the clean post-schema checkpoint ($n = 300$) and the
complete baseline validation run ($n = 2{,}195$) confirmed large-scale revision generalization.

\begin{table}[H]
\centering
\caption{Pre/Post Instrument Comparison (H5): Dimension-Level Results}
\label{tab:prepost}
\small
\resizebox{\textwidth}{!}{%
\begin{tabular}{p{3.5cm}cccccp{4cm}}
\toprule
\textbf{Dimension} & \textbf{Pre M} & \textbf{Post M} & \textbf{$\Delta$} & \textbf{Cohen's $d$} & \textbf{$p$} & \textbf{Interpretation} \\
\midrule
Tool Execution            & 3.020 & 4.775 & \textbf{+1.755} & \textbf{3.247} & $<.001$ & Dominant effect --- grounded revision confirmed at scale \\
Knowledge Transfer        & 3.663 & 4.008 &         +0.344  &         0.444  & $<.001$ & Grounded revision generalized beyond targeted cohort \\
Complexity Handling       & ---   & ---   &         +0.148  &         0.308  & $<.001$ & Small positive; no targeted revision \\
Cognitive Load            & ---   & ---   &         $-0.193$  &         0.388  & $<.001$ & Small negative --- sample composition shift \\
Memory Integration        & ---   & ---   & ---             &         0.131  & .034    & Negligible --- functionally stable \\
Content Quality           & ---   & ---   & ---             &         0.117  & .057 (ns) & Negligible --- stable across datasets \\
Hallucination Control     & ---   & ---   & ---             &         0.116  & .060 (ns) & Negligible --- stable across datasets \\
Processing Efficiency     & ---   & ---   & ---             &         0.095  & .123 (ns) & Negligible --- stable across datasets \\
Process-Outcome Alignment & 4.820 & 4.474 &         $-0.346$  &         0.676  & $<.001$ & Ceiling-inflation artifact in pre-checkpoint \\
\midrule
\textbf{Overall MAAC}     & \textbf{4.137} & \textbf{4.316} & \textbf{+0.179} & \textbf{0.615} & \textbf{$<.001$} & Medium overall --- dominated by TE revision \\
\bottomrule
\end{tabular}}
\tablenote{Pre = clean post-schema checkpoint ($n = 300$); Post = full baseline validation run ($n = 2{,}195$). Cohen's $d$ reported as absolute magnitude; direction encoded in $\Delta$ column. Tool Execution effect size ($d = 3.247$) is approximately $5\times$ larger than any other dimension.}
\end{table}

\subsection{Readiness Assessment}
\label{sec:readiness}

After seven iterative revision cycles culminating in v4.7, the validation program yields the
following readiness assessment.

\textbf{Tier 1 criteria:}
\begin{itemize}[noitemsep]
  \item All 9 $\alpha \geq 0.70$: \checkmark\ Met (v4.7 range: 0.770--0.922)
  \item All 54 $\lambda \geq 0.40$: $\times$ Not met --- 1 item below threshold (HC~Q6 $= 0.315$)
  \item All CFA converge: \checkmark\ Met (all 9 dimensions converge successfully)
\end{itemize}

\textbf{Tier 2 criteria:}
\begin{itemize}[noitemsep]
  \item CFI $\geq 0.90$ for all dimensions: $\times$ 8/9 pass (CQ remains below threshold;
        4/9 achieve Good fit)
  \item RMSEA $\leq 0.08$: $\times$ 0/9 pass, but frameable per \citet{Kenny2015} given
        $df = 9$ at $n > 2{,}000$
\end{itemize}

\textbf{Hypothesis support:} H1~(structural coherence)---strong support; all 9 dimensions
$\alpha \geq 0.70$ and 8/9 CFI $\geq 0.90$. H2~(differential quality)---strongly confirmed.
H3~(condition-awareness sensitivity)---partial support through the TE revision program;
final matched comparison remains future work. H4~(trace diagnostic value)---supported.
H5~(revision stability)---supported; v4.7 changes produced negligible collateral effects.

\textbf{Readiness judgment:} The six-component MAAC survey format is validated as a
fit-for-purpose measurement architecture. 8/9 dimensions meet Acceptable or Good CFA fit;
all 9 meet $\alpha \geq 0.70$. HC~Q6 remains the sole below-threshold item ($\lambda = 0.315$),
and the instrument is ready for baseline empirical use with documented caveats.

% -------------------------------------------------------
% 5. DISCUSSION
% -------------------------------------------------------
\section{Discussion}
\label{sec:discussion}

\subsection{What This Study Established About the Assessment Method}
\label{sec:established}

This study established five things not known after prior framework development alone. First,
the MAAC adjudication layer is a real measurement instrument that successfully surfaces
construct-operationalization quality. Second, the repeated six-component architecture is
methodologically worthwhile even when a dimension is not yet fully validated. Third, targeted
grounding revision works, generalizes, and scales. Fourth, the assessment method is sensitive
to complexity structure ($F(2, 591) = 114.67$, $\eta^{2} = 0.280$). Fifth, AI-administered
instruments can be developed iteratively using classical measurement-science principles.

\subsection{Why Schema Stability and Construct Validity Had To Be Separated}
\label{sec:separation}

A well-formed JSON schema does not imply a valid instrument; it only implies that the instrument
is being emitted consistently. The present study had to solve both problems in sequence. Schema
enforcement was a prerequisite but insufficient condition for psychometric validation.

\subsection{The Revision Playbook and Its Generalization}
\label{sec:playbook}

The two proof-of-concept revision cases illustrated different classes of assessor-instrument
failure. Tool Execution was a \textbf{condition-awareness failure}: the back half implicitly
assumed more tool use was better. Knowledge Transfer was a \textbf{proxy-formula failure}: the
back half used precise-looking ratios whose numerators were not grounded in stable observable
units. The same general revision playbook worked in both cases: preserve the stronger front half
where possible; isolate the weak back half; define observable units first; then revise the prompt
around those grounded units.

\begin{figure}[H]
  \centering
  \includegraphics[width=0.95\textwidth]{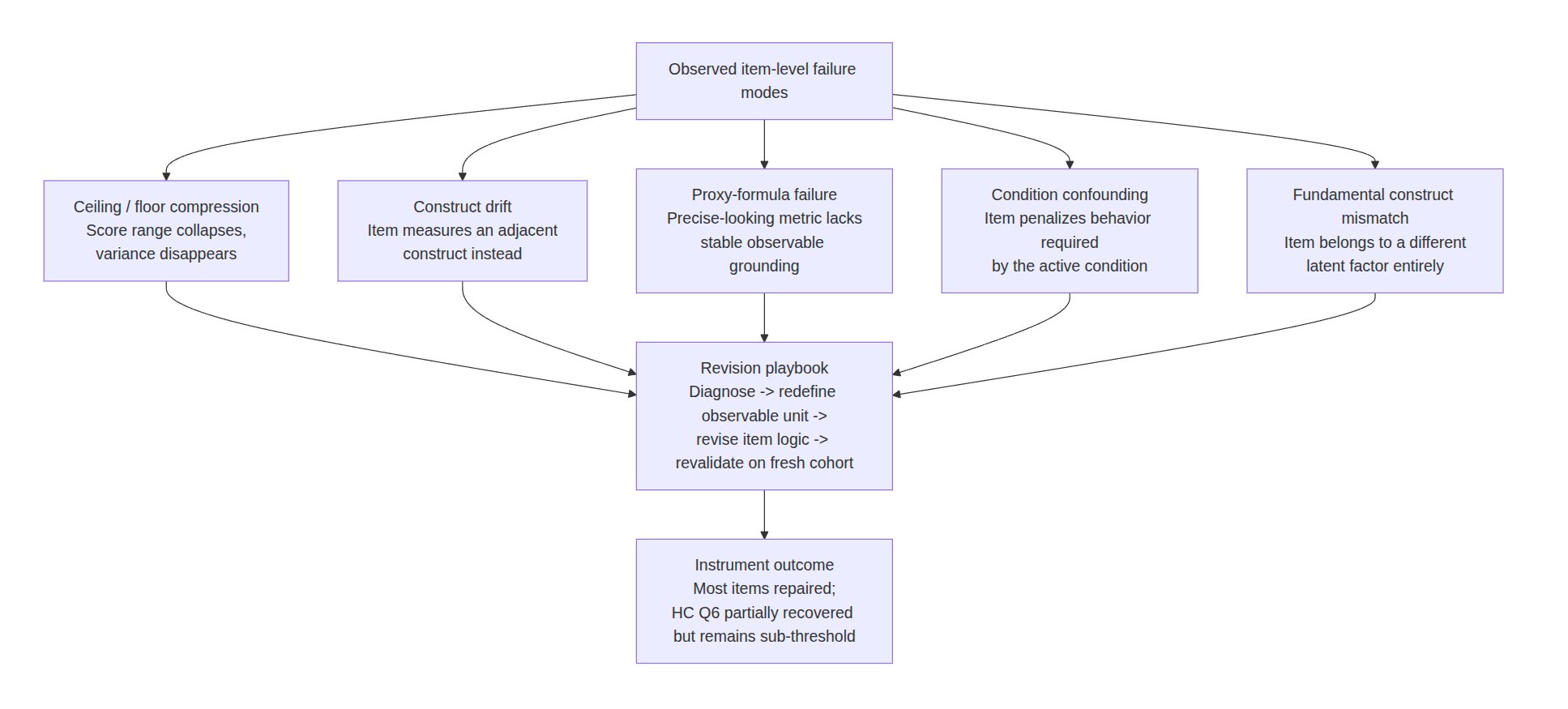}
  \caption{Five-category taxonomy of item-level failure modes in AI-administered instruments,
    with the shared revision playbook and instrument outcome. Observed failure modes feed into a
    common revision workflow (diagnose $\to$ redefine observable unit $\to$ revise item logic $\to$
    revalidate on fresh cohort), yielding an instrument where most items are repaired and HC~Q6
    is partially recovered but remains sub-threshold.}
  \label{fig:taxonomy}
\end{figure}

Figure~\ref{fig:taxonomy} shows the complete five-mode failure taxonomy and revision playbook
that emerged from the iterative program.

The five failure modes represent the complete space of construct-operationalization failures
encountered in AI evaluation instruments: (1)~\textbf{ceiling/floor compression} (PE~Q4);
(2)~\textbf{construct drift} (CL as a whole); (3)~\textbf{proxy-formula failure} (KT~Q5/Q6);
(4)~\textbf{condition confounding} (TE); and (5)~\textbf{fundamental construct mismatch}
(HC~Q6). Different failure modes require different revision strategies, but the diagnostic
playbook --- CFA-driven flagging, root-cause analysis, grounded item replacement, fresh-cohort
revalidation --- applies across all five.

\subsection{Current Readiness State and Remaining Gaps}
\label{sec:gaps}

Three bounded gaps remain. HC~Q6 remains the principal unresolved component, but the v4.7
\texttt{evidential\_calibration} redesign recovered the item to $\lambda = 0.315$---the best
loading achieved for this slot to date---making it partial construct recovery rather than a
confirmed permanent mismatch. The condition-sensitive comparison remains incomplete: Tool
Execution was successfully repaired as a baseline construct, but the final matched no-tools
versus tools-enabled validation remains future work. Content Quality remains below the 0.90
CFI threshold despite acceptable reliability and all items loading above threshold.

\subsection{Limitations}
\label{sec:limitations}

Five limitations require acknowledgment. First, the unit of analysis is an adjudicated survey
instance, not an independent human rater or a large population of distinct AI models.
Psychometric claims should be interpreted as validation of the instrument under an AI
adjudication regime. Second, the iterative revision program used fresh-cohort validation at each
cycle rather than a single frozen panel. Third, the five failure-type taxonomy emerged
inductively rather than being specified a priori. Fourth, the final matched tools-enabled
comparison was not completed in this paper. Fifth, Content Quality remains below the
CFI $\geq 0.90$ benchmark despite acceptable reliability.

\subsection{Implications for Empirical Application}
\label{sec:implications}

Subsequent empirical work can proceed with confidence that the MAAC instrument meets the
psychometric prerequisites for baseline empirical application. The current best v4.7 instrument
provides: reliable dimension scores (all 9 $\alpha \geq 0.70$); valid factor structure
(8/9 CFI $\geq 0.90$, 53/54 $\lambda \geq 0.40$); tier sensitivity ($\eta^{2} = 0.280$); and
a documented construct-revision trail supporting transparency about each dimension's
operationalization. The main caveats are HC~Q6 (retained with documented caveat) and the
deferred tools-enabled condition comparison.

% -------------------------------------------------------
% 6. CONCLUSION
% -------------------------------------------------------
\section{Conclusion}
\label{sec:conclusion}

This study asked a single organizing question: \textit{Is the MAAC assessor survey fit for
purpose?} The iterative validation program answers it with qualified confidence: the instrument
is publication-ready, with one documented sub-threshold item and one deferred condition-comparison
task reserved for future work.

Over seven iterative revision cycles spanning instrument versions v4.0 through v4.7, items were
revised using a disciplined CFA-diagnostic $\to$ root-cause $\to$ revision $\to$ revalidation
methodology. The results are substantive: all nine dimensions achieve $\alpha \geq 0.70$, 53 of
54 items load $\geq 0.40$ on their parent factor, and 8 of 9 dimensions meet Acceptable or Good
CFA fit (CFI $\geq 0.90$). The instrument differentiates meaningfully across complexity tiers
($\eta^{2} = 0.280$), confirming that MAAC scores are sensitive to the task-complexity gradient
built into the validation scenario pool.

The sole remaining below-threshold item is HC~Q6. The v4.7 redesign improved this item from
the failed v4.6 result ($-0.074$) to $\lambda = 0.315$---the best loading achieved for this
slot---but still below the target threshold. The instrument therefore ends this paper in a
strong but not fully closed state: broadly validated, with one partially recovered component
and one deferred tools-enabled comparison reserved for future work.

The most important methodological contribution is the demonstration that AI-administered
evaluation instruments can be developed iteratively using classical measurement-science
principles. The five failure-type taxonomy (ceiling/floor compression, construct drift,
proxy-formula failure, condition confounding, fundamental construct mismatch) and the associated
revision playbook provide a reusable framework for the emerging field of AI evaluation
instrumentation.

% -------------------------------------------------------
% STATEMENTS AND DECLARATIONS
% -------------------------------------------------------
\section*{Statements and Declarations}

\textbf{Competing Interests:} The authors declare no competing interests, financial or
non-financial, that are directly or indirectly related to this work.

\textbf{Funding:} No funding was received for this research.

\textbf{AI Tools Disclosure:} AI language model tools were used in editing and refining language
in this manuscript. All core contributions, analyses, interpretations, and conclusions are those
of the authors. The authors take full responsibility for the accuracy and integrity of all content.

% -------------------------------------------------------
% DATA AVAILABILITY
% -------------------------------------------------------
\section*{Data Availability}

The scenario corpus and hypothesis testing datasets from Study~1 that underpin this instrument
validation are publicly available via Zenodo:

\begin{itemize}[noitemsep]
  \item MAAC Complexity-Validated Decision Scenario Corpus v1.0:\\
        \url{https://doi.org/10.5281/zenodo.19776346}
  \item MAAC Study~1 Hypothesis Testing Dataset:\\
        \url{https://doi.org/10.5281/zenodo.19776734}
\end{itemize}

The instrument validation dataset generated in this study (adjudication traces, dimension scores,
and revision cohort data) is archived at:\\
\url{https://doi.org/10.5281/zenodo.19776955}

The preprint is archived at:\\
\url{https://doi.org/10.5281/zenodo.20114584}

% -------------------------------------------------------
% REFERENCES
% -------------------------------------------------------
\bibliographystyle{apalike}
\bibliography{references}

\end{document}